# Cyber Insurance, Audit, and Policy: Review, Analysis and Recommendations

Danielle Jean Hanson
Department of Computer Science
North Dakota State University
1320 Albrecht Blvd., Room 258
Fargo, ND 58108
Phone: +1-701-231-8562
Fax: +1-701-231-8255
Email: danielle.jean.hanson@ndsu.edu

Jeremy Straub[1]
Center for Cybersecurity and AI
University of West Florida
220 W. Garden St., Suite 250
Pensacola, FL 32502
Phone: +1-850-474-2999
Email: jstraub@uwf.edu

**Abstract**

Cyber insurance, which protects insured organizations against financial losses from cyberattacks and data breaches, can be difficult and expensive to obtain for many organizations. These difficulties stem from insurers difficulty in understanding and accurately assessing the risks that they are undertaking. Cybersecurity audits, which are already implemented in many organizations for compliance and other purposes, present a potential solution to this challenge. This paper provides a structured review and analysis of prior work in this area, analysis of the challenges and potential benefits that cyber audits provide and recommendations for the use of cyber audits to reduce cyber insurance costs and improve its availability.

**Keywords:** cyber insurance, cyber audit, risk-based cyber audit, cyber insurance pricing

## 1. Introduction

The growing number of cyber incidents and their economic impact is driving a need for – and driving up the cost of – cyber insurance. The average cost of a data breach in the United States was $10.22 million in 2024 [1]. Cyber insurance allows organizations to transfer some of the financial risk of cyber-attacks to a third party [2]. The National Association of Insurance Commissioners (NAIC) reported 38,496 cyber insurance claims were closed in 2024, with or without payment [3].

To profit from taking on the financial risk of cyber insurance without pricing out too many customers, insurers must conduct actuarial risk assessments of insured – and prospectively insured – organizations. Depending on the insurer, the policy, and the organization, the information for these risk assessments may come from industry and national data and statistics, organization-completed checklists, security assessments, and cyber audit reports [4].

---

[1] This work was partially completed while J. Straub was at North Dakota State University.

A cyber audit is an independent assessment, leading to an opinion, of an organization's cybersecurity controls. Cyber audits have scopes ranging from assessing a single device to an entire organization. Compliance audits check the audited system's conformance with a standard, such as NIST SP 800-171 [5], or a law, such as the Health Insurance Portability and Accountability Act (HIPAA) [6]. These audits are common in heavily regulated industries, such as healthcare and banking. Insurers can also require compliance-based cybersecurity audits by providing a checklist of controls that auditors can attest to an insured – or prospectively insured – organization's compliance with.

Unlike the checklist approach, risk based audits use procedures tailored to the organization's specific risk profile [7]. This requires an initial risk assessment of the organization, but it can result in more efficient audit procedures. It allows the audit to be customized and focus on what is of the greatest concern for the auditee.

This paper presents a review and analysis of the impact of cyber audits on – and the benefits of cyber audits for – cyber insurance. It continues, in Section 2, by providing background on cyber insurance, cyber audit, and cyber policy. This is followed, in Section 3, by a discussion of the relationship between cyber insurance, cyber audit, and cyber policy. Then, recommendations for insurers and organizations are presented. Next, the paper presents an analysis of how cyber policy can influence cyber insurance, before concluding and discussing areas of potential future work.

## 2. Background

This section provides background information regarding cyber audit, insurance and policy. The analysis and recommendations in later sections build on this prior work.

### 2.1. Risk Assessment

Insurers can prepare for underwriting losses by reserving cash in preparation of claims and by reinsuring against losses [8]. Insurers can also prepare for underwriting losses through risk assessment. Unlike reserving and reinsuring, risk assessment is proactive in reducing underwriting losses by allowing insurers to set appropriate prices and refuse customers that are likely to be unprofitable.

Risk assessment can also be a tool to encourage security controls among organizations in the industries and the regions where cyber insurers operate by allowing organizations which have reduced their cyber risk to receive preferential pricing. Cyber insurance risk assessment looks at organizational loss history, size, industry, security controls, at-risk assets, security assessment reports, and other information to determine both the amount at risk, if a cyber event occurs, and the likelihood of a cyber event occurring. Risk assessment requires upfront, reliable, accurate information about the state of the organization [9]

Because cyber insurance is a relatively immature industry, and cyber threats are rapidly involving, cyber insurers must frequently issue new policy versions [10]. Cyber insurers must also consider the risk of a large-scale cyber-attack, such as the 2017 NotPetya malware attack [11], which might simultaneously affect multiple or even all policyholders. An event like this puts a strain on insurer resources and reinsurance providers [12].

According to the NAIC, nearly 60% of the $16.6 billion spent on cyber insurance premiums in 2023 was spent in the United States [13]. The following year, $3.9 billion was spent on cyber insurance premiums in Europe [14]. Micro-insurance and personal cyber-insurance have potential in developing markets, such as Nigeria, where residents are online and engaging in a digital economy [15], but where businesses have greater barriers to organizational cyber insurance policies than businesses in wealthier or more regulated countries, such as the United States.

Personal cyber-insurance allows individuals and families to guard against cyber risk that an organization may not have insurance or resources to cover. This provides protection when engaging with vendors and other organizations that do not have adequate cybersecurity controls in place [16].

### 2.2. Cyber Attack Costs

Cyber attacks and data breaches have many direct and indirect financial costs to their targets. These result from fines and legal fees that may need to be paid, due to a data breach, the cost of remediation after an incident, the cost of lost business due to downtime and reputational damage. An organization may also incur the cost of lost productivity during a business interruption and the loss of competitive advantage, as a result of corporate espionage and theft of trade secrets.

When cyber-attacks lead to physical effects, there may be a cost of wasted materials, environmental cleanup, and even human safety and life related costs [17, 18]. Small and medium enterprises (SMEs) are frequent targets of potentially crippling cyber-attacks; however, they can often obtain cyber insurance at lower rates than large companies. This is for several reasons including policy requirements and pricing.

### 2.3. Cyber Audit

There are several types of cyber audits and also several adjacent types of audits. Two common types of audits, that are inclusive of cybersecurity controls, are Systems and Organization Controls (SOC) 2 audits and integrated audits. SOC 2 audits are designed to provide service organizations with a way to demonstrate their security controls to customers and other business affiliates [19]. Integrated audits, which are required by the Sarbanes-Oxley Act of 2002 (SOX) [20], involve an audit of the internal controls over financial reporting which, for most organizations, includes certain cybersecurity controls [20].

The two major barriers to cyber auditing are time and financial cost. There can also be tension between audit effectiveness and audit efficiency [21]. The more testing an audit includes, the more time and resources that are required. While automation, effective planning and risk assessment can improve audit efficiency, it is not feasible for an audit to provide total assurance [22].

Other barriers to cyber audit effectiveness are organizational and employee resistance. Organizations may not see a value in dedicating resources to undergoing a cyber audit. Audits take time away from employees' ordinary work duties, as they must provide documentation to auditors and attend audit meetings and interviews. Individuals within an organization may also be resistant to what they may consider criticism of or attempts to find fault with their work.

### 2.4. Cyber Policy

Nations, states, provinces, and local governments have policies, as do organizations. Policy includes and is shaped by local, national, and international law. It is also shaped by organizational culture, risk tolerance, association and industry standards, customer and affiliate expectations, and more.

Examples of laws that demonstrate or affect cybersecurity-related policy include data privacy laws like the Health Insurance Portability and Accountability Act (HIPAA) [6], the Gramm-Leach-Bliley Act

(GLBA) [23], the Family Educational Rights and Privacy Act (FERPA) [24], the General Data Protection Regulation (GDPR) [25], and the Children's Online Privacy Protection Act (COPPA) [26].

Many federal contractors are required to meet certain security standards outlined in the Cybersecurity Maturity Model Certification (CMMC) requirements [27]. Laws, such as SOX [20], which regulate reporting can also affect cyber policy. SOX requires an audit of and reporting on the internal controls over financial reporting of publicly traded companies in the United States. These internal controls include information technology controls.

In addition to statutes, policy also comes from governments in softer forms. This includes recommendations like the UK Code of Practice for Consumer IoT Security [28], which imposes requirements on contractors, or provide resources to individuals and organizations. Associations and other non-government organizations also implement policies. An example of this is Payment Card Industry Data Security Standard (PCI DSS) [29], which banks require merchants that directly process credit card transactions to comply with. Associations may also impose requirements upon their members to preserve industry reputation.

Nations have different cybersecurity policy focuses and approaches [30], as do organizations. There is a tension between collaboration and competitiveness among nations and organizations, as there is a measure of vulnerability in sharing information about cyber-attacks and challenges with outsiders. Because of this, not all nations and organizations are equally collaborative in their cyber policy. Regardless of what the particular policies are within an environment, they influence cyber audit and cyber insurance pricing and availability.

**3. Cyber Audit and Cyber Insurance Pricing**

This section discusses the relationship between cyber audit and cyber insurance pricing. It also reviews how applicable standards, legislation, and audit techniques can affect cyber insurance.

*3.1. Cyber Audit and Cyber Insurance*

Cyber insurers have a variety of pricing models to choose from when pricing cyber insurance for their customers [31]. Regardless of which pricing model they choose, cyber insurers benefit from an accurate risk assessment of the organizations they insure when setting pricing.

Problematically, cyber-insurance pricing has been found to not adequately reflect cyber-insurance risks [32]. Simply gathering general, readily confirmable information about an organization's size, industry, and geographic distribution does not provide a comprehensive picture of its cybersecurity risk.

Insurers can gather more detailed information about an organization's cybersecurity posture using annual security control checklists and reporting. However, checklist-based reporting cannot be equally effective for all organizations without becoming complicated and unwieldy. Additionally, without a mechanism for insurers to independently verify the insured organization's self-reported control checklist, the insurer must rely entirely on the insured organization's self-reporting, which may be inaccurate. Although policies can exclude coverage for incidents resulting from the insured organization's failure to maintain the required control standards, it is better to reliably identify these issues before a claim occurs [33].

An assessment conducted either by or on behalf of the insurer or by an independent auditor can provide assurance regarding the appropriateness of an insured entity's risk assessment. Additionally, some pricing models require insured organizations to undergo an audit in order to be in effect at all [34].

Audits can be very costly. While highly regulated organizations, such as publicly traded companies [20] and financial institutions [24], may already be required to undergo independent audits of internal controls that can include cybersecurity controls, small or lightly regulated organizations may not already be subject to regular audits of their cybersecurity controls. Requiring audits for cyber insurance can, thus, place a significant burden on those organizations. Insurance companies must balance the risk of miscalculating insurance risk with the risk of driving away potential customers through insurmountable financial barriers to coverage.

Risk-based auditing may present a solution to this challenge. Through risk-based auditing, that tailors cybersecurity audits to the individual auditee, and through advances in cybersecurity audit automation — as well as improved management responses to audits – it may be possible to improve audit effectiveness and efficiency. Lowering the cost improves the value proposition that audits provide to auditees and increases audit accessibility. Improved cybersecurity audits would improve insurance providers risk assessments and allow them to adjust pricing models accordingly, which could decrease the cost of insurance for some organizations.

Cyber insurers are interested in both the incident preparation and the incident response plans of their clients [35]. This corresponds to preventive and corrective controls. As such, a cybersecurity audit for the purpose of obtaining insurance coverage should not neglect the corrective controls and incident response and business continuity plans of the auditee.

Aziz, Suhardi and Kurnia [10] identified the high cost of external audits as a challenge to insurers when setting contractual requirements for the insured. The insurer must believe the additional information and peace of mind of not relying solely on the insured's self-reporting is worth the financial cost, to justify requiring the audit.

### *3.2. Cyber Insurance Pricing and External Audit*

Cyber insurers must balance insurance risk – including the potential of a major cybersecurity incident affecting multiple clients simultaneously – with the impact of setting cyber insurance pricing too high for it to be a viable product. Considering that only a minority of SMEs have cyber insurance, increased barriers to cyber insurance access could lead to diminished clients and revenue for cyber insurers [36].

Concerns about cyber-insurance pricing are not limited to SMEs in the United States. Insurers in other countries like Greece – which only had one local cyber-insurer in 2025 – and Cyprus, also struggle with building appropriate actuarial models for cyber-risk [37]. This can lead to caution by insurers when entering the cyber insurance market and limit competition and availability in the cyber insurance industry.

Emerging economies are also heavily impacted by cybersecurity challenges [38]. Businesses around the world struggle with cyber-insurance access and affordability. A 2025 survey of 241 Nigerian organizations found that 38.2% were not aware of cyber insurance, and most SMEs were unsure about or uninterested in cyber-insurance. This was primarily because of a lack of cyber-insurance affordability and organizational readiness. This is despite Nigeria being the nation most frequently targeted by cyber-attacks in Africa [39].

### 4. Structured Literature Review and Analysis

To better understand the relationship between cyber insurance pricing and external audits, existing work in the area was reviewed. A search was performed with Google Scholar using the keywords "cyber insurance" AND "external audit" AND "cost of" AND "premiums" on November 11, 2025. Of these results, only English-language articles focusing on cyber insurance or external audit as primary topics were considered. This left only a handful of relevant articles.

One was from 2015 and one was from 2017. Two were published in 2024 and one was from 2025.

"Designing Model for Calculating the Amount of Cyber Risk Insurance" [34] (2017) presented a cyber insurance pricing model that takes into account the customer's compliance to standards and the results of internal and external audits of the customer.

"The Role of Cyber-Insurance, Market Forces, Tort and Regulation in the Cyber-Security of Safety-Critical Industries" [40] (2015) discussed both cyber-insurance and government, along with market forces and torts, as forces for responding to cybersecurity concerns. One of the cyber-insurance challenges mentioned within the article is the limited actuarial data for building pricing models.

"Cyber insurance risk analysis framework considerations," [4] (2024) surveyed cyber insurance professionals to learn about the challenges of and considerations regarding risk analysis for cyber insurance pricing. This study found that cyber insurers are not equally able to perform risk assessments of insured companies, and those insurers that are able to perform risk assessments do not share a universal approach.

"Cybersecurity Risk and Audit Pricing—A Machine Learning-Based Analysis," [41] (2024) presented a machine learning algorithm that considers several metrics, including external audit.

"A Meta-Analysis of Cybersecurity Framework Integration in GRC Platforms: Evidence from U.S. Enterprise Audits," [42] (2025) found that integrating cybersecurity frameworks into government, risk, and compliance (GRC) systems of certain highly regulated industries can significantly improve cybersecurity outcomes and audit readiness. However, the meta-analysis also showed that many enterprises and sectors find challenges in implementing cybersecurity frameworks.

Despite the few papers surveyed, two themes were of note. One of them was the challenge of building pricing models [34, 40, 41]. The second notable theme, among the surveyed papers, is that there are ways to address the challenges related to cyber insurance costs.

With improved auditing frameworks and practices and potential legislative action, a pricing model that uses audit results and standard compliance, such as that introduced in [34], could potentially solve the pricing challenge for cyber insurers. A challenge faced when using standards and regulation as a tool for managing cyber risk is that security standards designed to be broadly applicable do not always fit individual industries and organizations in the same way. This challenge can be addressed by using risk-based auditing together with compliance audits to more accurately assess cyber risk and allow insurers to refine pricing models to consider the broader organizational picture of insured organizations.

## 5. Solution Components

This section discusses potential components to a solution for cyber insurance pricing and availability. Section 5.1 discusses technology and audits. Section 5.2 presents analysis regarding cyber insurance audit standards as well as on relevant legislation and techniques.

### *5.1. Technology and Audits*

Technology, such as machine learning [41], can be used to aid insurers when tailoring risk assessments to specific insured organizations. The use of risk analysis frameworks can promote consistency in risk assessment methodology and pricing determination and improve transparency in cyber insurance pricing [4]. Finally, cybersecurity frameworks – put into place by organizational governance, risk, and compliance systems – can improve the audit process for insured organizations who undergo external audits [42].

External audits are a valuable tool for determining risk and pricing in the cyber insurance industry. However, their expense is a barrier to their widespread use, so insurers frequently rely on insured self-reporting. Additionally, different insurers use different risk assessment methods and pricing models. Particularly in cases such as businesses with large IoT networks, audit automation and continuous auditing tools have the potential to bring auditing costs down significantly [43].

Risk-based cyber audits, that follow a standardized framework, can be used to improve pricing models and heighten comparability of audit results among organizations and industries, which would decrease risk to insurers. This decreased risk to the insurer could then benefit the insured in the form of lower premiums. Together with improvements in automation, a standardized, risk-based cybersecurity auditing framework could improve the cost-benefit balance of requiring insured parties to undergo an external audit before being eligible for cyber insurance or discounted rates.

*5.2. Cyber Insurance and Cyber Security Audit Standards, Legislation, and Techniques*

There are a variety of cybersecurity standards that exist and which are commonly used as a baseline for evaluating organizational security, even in organizations where adherence to these standards is not required by regulation. Examples include the NIST Protecting Controlled Unclassified Information in Nonfederal Systems and Organizations (NIST SP 800-171) standard [5] and the International Organization for Standardization (commonly known as the ISO) standard Information Security, Cybersecurity and Privacy Protection — Information Security Management Systems — Requirements (ISO/IEC 27001) standard [44].

Some standards, such as the National Institute of Standards and Technology Considerations for Managing Internet of Things (IoT) Cybersecurity and Privacy Risks (NIST IR 8228) [45], the European Union Agency for Cybersecurity (ENISA) Baseline Security Recommendations for IoT [46], and the European Telecommunications Standards Institute Consumer IoT Security Road Map (ETSI EN 303 645) [47] can be applied specifically to IoT devices and networks. These standards can be used as a baseline for creating cyber insurance internal control checklists. Compliance with these standards can be considered when building cyber insurance pricing models.

Relevant legislation for insured organizations is also a good starting point for evaluating the insurance risk of an organization. For example, insurance companies may require U.S. healthcare providers to provide evidence of HIPAA compliance [6] or U.S. financial institutions to provide evidence of Gramm-Leach-Bliley Act (GLBA) compliance [23]. In many cases where an insured organization operates within a heavily-regulated environment, the organization already undergoes external audits or other assessments that can be referenced by the insurer to prevent duplicative work.

There are many ways to assess organizational security controls. Self-reporting, compliance reviews, SOC audits, internal audits, and security control checklists are just a few examples. An IT security effectiveness evaluation model that uses financial spending as a benchmark [48] has also been proposed.

Only an independent audit, however, provides formal assurance as to the organizational controls within the scope of the assessment. The financial and time costs of audits increase along with the scope of the audit, and reasonable assurance – the highest level of assurance auditors provide – is still not absolute assurance. Security standards and legislation provide a baseline for compliance audits, which verify the organization is compliant with the requirements of the relevant standard or legislation, which makes compliance audits a good tool for verifying insured organization self-reported controls are in place.

Risk-based audits, which have audit procedures tailored to each organization's risk assessment, are useful tools for when a standard control checklist may be inadequate or inapplicable for some aspects of the insured organization's security. Whether an audit is risk-based or compliance-based, it can benefit from automation tools and techniques that can efficiently examine entire networks rather than relying on samples like traditional audit techniques do.

**6. Cyber Audit Recommendations**

This section presents recommendations for cyber audits, based on the analysis presented in previous sections. Notably, there is limited prior work on this specific topic, and these recommendations seek to fill the existing gaps.

As much as possible, insurers should utilize results from audits and assessments insured – and prospectively insured – organizations already undergo. Security control checklists for insured organizations to follow should be tailored, as much as possible, to organization industry and level of technological adoption. For high-risk organizations that are unable to provide sufficient documentation from other audits, insurers should require regular audits of internal cybersecurity controls. When it is not practical or cost-effective to require regular audits of internal controls, without losing business, an option is to require audits of a subset of internal controls or systems every year or audit the organization at intervals greater than one year.

Risk-based audits of internal controls should be utilized when ordinary checklists and standards are insufficient or largely inapplicable for an organization, such as in manufacturing where much of the cybersecurity risk comes from operational security risks rather than information security risks. Cyber insurers may benefit from providing incentives, such as reduced rates, for organizations which voluntarily implement practices such as continuous audit automation or other security monitoring measures.

Insured organizations should research steps that they can take to lower their cyber insurance premiums. They should also see if there are ways to communicate or structure security policies that they already implement so that their insurer can consider them when setting premiums, as these may lower the insured organization's assessed cybersecurity risk.

If organizations are required to undergo a security audit to obtain or to afford cyber insurance coverage, they should try to derive as much benefit from the audit as possible. This can be done by using audit findings and recommendations as tools for identifying areas of cybersecurity improvement. Internal audit and other internal assessment activities should be structured to take advantage of the external audit and avoid duplicative procedures.

One answer to cyber insurance unaffordability among SMEs is creating legislation that shifts some burden to the individual consumer. This would protect SMEs and also deal more effectively with the fact that many personal data misuse incidents cannot be traced to a specific data breach. Individuals could purchase personal insurance and micro-insurance. Personal cyber insurance is purchased by individuals and households to protect against their cyber risk [16]. This transfers the financial burden of insurance more directly to the consumer.

Micro-insurance is a form of personal cyber insurance that is specifically designed for very low-income households [15]. It allows low-income individuals in developing countries, who cannot afford many traditional global cyber-insurance offerings, to afford cyber insurance.

## 7. Cyber Policy Analysis

The United States does not require organizations to purchase cyber insurance in the way state governments require drivers to purchase auto insurance (see, e.g., [49]). Because of this, organizations are price sensitive when choosing whether to purchase cyber insurance or not and must be convinced that the potential benefits of coverage outweigh the recurring premium costs.

There are policy approaches that governments can take that improve cyber insurance adoption and access in the United States without forcing it upon organizations. These include encouraging the adoption of risk-based audit strategies, as discussed in previous sections, and providing government reinsurance for cyber insurance [50], potentially lowering insurance costs.

Audited financial reports are consistent and comparable between auditors, because there have been efforts from governments (e.g., SOX [20]) and within the accounting profession to regulate and standardize the reporting process, while allowing room for the professional judgment of auditors [51]. Policy efforts from national and international standard-setting bodies – working with effected organizations and cyber auditors – to implement a widely adopted, risk-based cyber audit framework could improve cyber auditing. This would provide insurers and other stakeholders with higher-quality, helpful audit reports that allow them to compare similar organizations with one another. Organizations would have a better idea of what to expect during the audit process, causing less confusion. Auditors would have benchmarks and resources to facilitate more efficient audits. These standards can include provisions for upcoming and future developments in audit and risk assessment automation to encourage their adoption.

Organizational policy can also improve cyber insurance coverage. Organizations can implement internal policies requiring that major vendors and other affiliates have cyber insurance coverage or that the lack of coverage be evaluated in determining whether to purchase from or work with them.

Cyber audit has the potential to improve the accessibility and profitability of cyber insurance. By providing insurers more consistent and reliable data for risk assessment, cyber audit allows insurers to price policies more competitively for organizations that are able to demonstrate they have effective security controls in place. National cyber policy and professional standard-setting bodies can facilitate this by creating and implementing risk-based cyber audit and reporting standards that benefit stakeholders both internal and external to the audited organizations.

## 8. Conclusions and Future Work

This paper has provided a review of the relationship between cybersecurity audits and insurance and discussed the potential benefits that cyber audits could provide to the current challenge of cyber insurance cost and availability. Current methods are limited by assessment accuracy which results in insurers having limited or inaccurate pictures of their insured organizations' risk level.

Cyber insurers use a variety of pricing models and risk assessment methods. The accuracy of the cyber insurer's risk assessment is directly related to the insurer's ability to reduce its own insurance risk, set pricing that matches the insured organization's risk assessment, and maximize profitability. Cybersecurity audits can be excellent tools for cyber insurance risk assessment, especially for organizations which already undergo cyber audit. By understanding organizations' risks, insurers can set competitive premiums that accurately take into account the cost of their own risk.

The principal barrier to widespread cyber auditing for insurance purposes is cost. The use of audit standards and automation can help improve cybersecurity audit efficiency. Audit frameworks can help expedite audits and aid auditees in using audit reports as tools for strengthening their security postures.

Reducing audit costs and increasing their benefits can aid SME adoption, potentially reducing barriers for SMEs to obtain cyber audits. Improved cyber insurance risk assessments can also improve cyber insurance affordability for SMEs and profitability for insurance providers.

Needed future work in this area includes conducting a more detailed assessment of the practices of cyber insurers and reinsurers, as well as organizations that have and do not have cyber insurance. This paper has shown that there is an acute problem with cyber insurance affordability and a potential solution from conducting effective and low-cost cyber audits. Future work will need to further assess the effectiveness of potential techniques for enhancing cyber audits and the real-world benefits of these audits to insurers.

**Acknowledgements**

This work was partially supported by the NDSU Sheila and Robert Challey Institute for Global Innovation and Growth.